# RYTHMES D'ACTIVITE LOCOMOTRICE CHEZ DEUX INSECTES PARASITOIDES SYMPATRIQUES : *EUPELMUS ORIENTALIS* ET *EUPELMUS VUILLETI (Hyménoptère, Eupelmidae).*

A. Ndoutoume-Ndong[1], D. Rojas-Rousse[2], R. Allemand[3]

[1] Ecole Normale Supérieure de Libreville, B.P. 17009 Libreville – Gabon.- Tel.(241) 26 16 54 fax. (241) 73 31 61 - email : augustendoutoume@caramail.com

[2] Institut de Recherche sur la Biologie de l'Insecte, UPRESA-CNRS 6035, Faculté des Sciences, Parc de Grandmont, 37200 Tours.

[3] Laboratoire de Biométrie, Génétique et Biologie des populations de Lyon I – Université Claude Bernard – 43, Bd du 11 novembre 1918, 69622 Villeurbanne.

**Résumé :** A l'aide d'un système automatique d'analyse d'images, nous avons étudié la répartition temporelle de l'activité locomotrice de *E. orientalis* et *E. vuilleti* au cours de 24 heures, et sur plusieurs jours pour savoir si les rythmes d'activité de ces deux Eupelmidae jouent un rôle dans leurs interactions compétitives. L'analyse des rythmes d'activité locomotrice d'*E. orientalis* et *E. vuilleti* montre que l'activité locomotrice des deux espèces présente des variations cycliques journalières. Ces deux Eupelmidae ont des rythmes d'activité semblables. Les déplacements de ces parasitoïdes ont lieu essentiellement pendant la photophase. Mais l'activité d'*E. vuilleti* est plus précoce puisque les individus de cette espèce démarrent leur activité en moyenne 4 à 5 heures plus tôt que ceux d'*E. orientalis*. *E. vuilleti* commence les déplacements plusieurs heures avant l'éclairage alors que *E.* orientalis n'est actif qu'en présence de la lumière. Ce décalage de démarrage d'activité est donc un facteur permettant aux deux espèces concurrentes de minimiser leurs interactions pendant la période de cohabitation dans les greniers après les récoltes du niébé.

Mots clés : Eupelmidae, parasitoïdes, activité locomotrice, rythme, variations journalières, décalage d'activité, interactions, cohabitation.

# LOCOMOTOR ACTIVITY RHYTHMS IN TWO SYMPTRIC PARASITOID INSECTS : *EUPELMUS ORIENTALIS* AND *EUPELMUS VUILLETI (Hyménoptera, Eupelmudae).*

**Abstract :** With an automatic image analysis device, we studied the temporal distribution of the locomotor activity of *E orientalis* and *E vuilleti* during 24 hours, and over several days to know whether the activity rhythms of these two Eupelmidae play a role in their competitive interactions. The analysis of locomotor activity rhythms of *E. orientalis* and *E vuilleti* shows that the locomotor activity of both species presents daily cyclic variations. These two Eupelmidae have similar activity rhythms. Displacements of these parasitoïdes essentially take place during the photophase. But the activity of *E vuilleti* is earlier because the individuals of this species start their activity on average 4 to 5 hours earlier than those of *E orientalis*. *E vuilleti* begins displacements several hours before lighting whereas *E orientalis* is active only in the presence of the light. This shift of starting activity is thus a factor allowing these concurrent species to minimize their interactions during the cohabitation period in traditional granaries after the harvests of niébé.

# Résumé en Anglais:

The biological rhythms are observed in the great majority of alive beings in the expression of molecular, biochemical, physiological or behavioural phenomena. The occurrence of these phenomena is considered as an adaptation to cyclic variations of physical environment. In parasitoïde insects, the circadian rhythms of activity play an important role in the coexistence several species because females are in general unable to recognize an already parasitized host by another species.

This study carried out temporal distribution of locomotor activity of two hymenoptera Eupelimidae insects (*Eupelmus orientalis* and *Eupelmus vuilleti*). These hymenoptera are two close related species encountered in the fields and in stocks after harvest. Those sympatric species are parasitoïdes of bruchidae insects which are leguminous plant pest in tropical Africa. In the fields as in stocks of niebe (Vigna unguiculata), *E orientalis* and *E vuilleti* parasitize larvae and nymphs of bruchidae. The coexistence of two species exploiting the same resources is possible only if exists mutual recognition or temporal sharing of resources. This study carried out the existence of rhythms in locomotor activity and the temporal distribution of these parasitoïdes according to daily activity and this activity over several days.

Using an automatic system of images analysis, we studied the temporal distribution of the locomotor activity of *E orientalis* and *E vuilleti* during 24 hours, and over several days. It is an effective tool because allowing an automatic follow-up of several individuals the day like the night. The automation of the system was not done with precision detriment of the measurements one combined the quantification of insect trajectory to the temporal follow-up. The analysis of locomotor activity rhythms of *E orientalis* and *E vuilleti* shows that locomotor activity of both species presents daily cyclic variations. These two Eupelmidae have similar activity rhythms. Displacements of these parasitoïdes take place early during the photophase. But *E vuilleti* activity is early because of this species start their activity on average 4 to 5 hours before *E orientalis*. *E vuilleti* begins displacements several hours before lighting whereas *E orientalis* is active only in light presence. This starting shift of the activity is thus a factor making possible two concurrent species to minimize their interactions during the cohabitation period in the granaries after harvests.

In fields, these Eupelmidae exploit the same hosts, for it they can consider that the activity shift is a means of reducing the competition intensity. An other study on the competition between these two species showed that *E vuilleti* eggs deposited on hosts parasitized by *E orientalis* are likely weak to reach the adult stage. That probably constitutes a pressure which led *E vuilleti* to start its daily activity earlier in order to parasitize available healthy hosts before discovered by *E orientalis* females. The role of the activity shift in the coexistence of the sympatric species was evoked for several insects associations. In the case of these Eupelmidae it is probable that *E vuilleti* early activity is an effectiveness factor of hosts search making it possible to compensate lower competitive capacity. The study of search capacity of hosts by these Eupelmidae must be planned in situation of competition in order to understand better the role of the early activity of *E vuilleti*.

# INTRODUCTION

En réponse aux variations périodiques des facteurs de l'environnement (alternance jour/nuit et cycles associés), les êtres vivants ont développé des phases d'activité et de repos en fonction de leurs exigences physiologiques et écologiques. Les rythmes biologiques sont observés chez la grande majorité des êtres vivants dans l'expression de phénomènes moléculaires, biochimiques, physiologiques ou comportementaux. L'expression de ces phénomènes est considérée comme une adaptation aux variations cycliques de l'environnement physique [1].

Chez les insectes parasitoïdes, les rythmes circadiens d'activité jouent un rôle important dans la coexistence des espèces car les femelles sont en général incapables de reconnaître un hôte déjà parasité par une autre espèce [2, 3, 4, 5]. La plupart des travaux sur la coexistence d'espèces de parasitoïdes appliquent les mécanismes classiques : spécificité parasitaire [6, 7, 8, 9], partage des ressources [10, 11, 12,], partage temporel à l'échelle saisonnière [13], hétérogénéité de l'habitat [14], et toute une série de stratégies reproductives décrites dans de nombreuses associations [15, 16, 17].

Au cours de ce travail, nous avons étudié la répartition temporelle de l'activité locomotrice de deux insectes hyménoptères Eupelimidae (*Eupelmus orientalis et Eupelmus vuilleti*). Ce sont deux espèces sympatriques de parasitoïdes des insectes bruchidae qui sont des ravageurs de légumineuse en Afrique tropicale. Dans les champs comme dans les stocks de niébé (*Vigna unguiculata*), *Eupelmus orientalis et Eupelmus vuilleti* parasitent des larves et nymphes de bruches. La coexistence de deux espèces exploitant les mêmes ressources n'est possible que s'il y a reconnaissance mutuelle ou partage temporel de ressources. C'est pourquoi l'objet de cette étude est de mettre en évidence l'existence de rythmes d'activité locomotrice et la répartition temporelle de cette activité chez ces parasitoïdes à l'échelle journalière et sur plusieurs jours.

**1- Matériel et méthode de mesure**

   **a)- Matériel biologique**

Dès l'émergence, les mâles et les femelles parasitoïdes sont séparés avant de s'accoupler ; les mâles sont ensuite placés dans une boite de Pétri et les femelles dans une autre. Pour avoir des femelles accouplées, on laisse cinq femelles en présence de cinq mâles âgés de deux jours. Après un seul accouplement, les femelles sont à nouveau isolées dans les boîtes de Pétri et sont prêtes pour les expériences. La mesure des rythmes d'activité a été faite dans deux conditions climatiques : celles qui sont similaires aux conditions climatiques qu'on observe dans la localité d'origine (Niger) de ces espèces pendant la saison des pluies avant la récolte du niébé (32° : 22°C, 50 % : 80 % h. r., L. D. 12 : 12), puis dans les conditions semblables à celles qui règnent dans cette région pendant la saison sèche au début du stockage du niébé (25°: 15°C, 30 % : 60 % h. r., L. D. 12 : 12).

   **b)- Présentation du matériel de mesure d'activité et acquisition des données**
   **b.1- Système de mesure**

Les mesures de l'activité locomotrice ont été réalisées au moyen d'un système automatique d'analyse d'image du Laboratoire de Biométrie et de Génétique des populations de Lyon I.

Le système automatique d'analyse d'image est composé d'enceintes de mesure, une caméra, un micro-ordinateur, un moniteur vidéo et une imprimante. La caméra vidéo (Canon CI-20PR) se déplace dans un plan parallèle à celui des enceintes de mesure grâce à deux axes de translation dirigés par des moteurs (Socitec, FDL 603-370-47). Le fonctionnement du moteurs est commandé par le programme de mesure qui assure donc à la fois l'analyse d'images et les mouvements de la caméra. La liaison entre les moteurs et le micro-ordinateur se fait grâce à une carte d'interface (IF1 Socitec).

Les images prises par la caméra sont transmises à l'ordinateur au niveau d'une carte vidéo (SECAD) permettant leur numérisation. Les images numérisées sont ensuite analysées par le logiciel. Les données enregistrées à chaque mesure sont stockées dans un fichier dont l'analyse en fin d'expérience permet la représentation des variations temporelles. A tout instant, les opérations d'analyse d'images sont visualisables sur un moniteur de contrôle.

Les cellules de mesure sont éclairées en permanence par transparence par une source infra rouge ($\lambda > 730$ nm), longueurs d'ondes auxquelles les insectes ne sont pas sensibles [18]. L'objectif de la caméra étant muni d'un filtre ayant la même bande passante, la prise d'image est indépendante des variations de la "lumière du jour" simulée par 2 tubes fluorescents (lumière blanche, intensité 350-500 lux).

Ce système d'analyse d'image est performant puisqu'il permet le suivi automatique d'un grand nombre d'individus de jour comme de nuit. L'automatisation du système ne s'est pas faite au détriment de la précision des mesures puisqu'on a allié la quantification du trajet de l'insecte au suivi temporel [19]. Ceci permet de considérer plusieurs aspects de l'activité locomotrice, tant qualitatifs que quantitatifs. De plus le système est adaptable à plusieurs situations car il permet aussi bien l'étude de l'activité spontanée que celle de l'activité en présence d'hôte ou celle des rythmes d'émergence.

**c- Conditions expérimentales**

Pour mesurer l'activité, les insectes sont placées dans les cellules de mesure (diamètre = 2,8 cm, épaisseur = 1 cm) réalisées dans des plaques de métal maintenues entre deux plaques de verre par des pinces métalliques. Chaque cellule contient des gouttelettes de miel dilué en quantité suffisante pour permettre la nutrition d'un individu pendant plusieurs jours.

Afin de limiter les variations de température et d'humidité, et bien que la pièce d'étude soit climatisée, les enceintes de mesure sont placées dans un dispositif en verre [19]. Etant donné que la pièce d'étude n'est pas équipée d'un système de régulation automatique de la thermopériode et d'humidité relative, les mesures ont été faites dans des conditions constantes de 28° C ± 2° C, 75 % h. r. pour les parasitoïdes qui se sont développés à 32° : 22°C, 50 % : 80 % h. r., L. D. 12 : 12. Pour ceux qui se sont développés en conditions 25°: 15°C, 30 % : 60 % h. r., L. D. 12 : 12, les mesures ont été faites à 21° C ± 2° C, 75 % h. r.. Toutes les mesures sont faites sous une photopériode de L.D. 12 : 12. La photophase commence à 8 heures et s'arrête à 20 heures, ainsi la scotophase dure de 20 heures à 8 heures.

### d- Paramètres calculés

- **Activité journalière** : c'est le pourcentage de temps consacré au déplacement en 24 heures. Il peut s'agir d'une moyenne sur plusieurs jours si on travaille sur le jour moyen.

- **Heure médiane d'activité (H. M. A.)** : c'est le moment de la journée auquel 50 % de l'activité journalière est réalisée [20]. Ce paramètre permet de distinguer les individus selon que leur activité est plus ou moins précoce.

- **Indice de profil (Ip)** : Cet indice permet de décrire la forme de la courbe d'activité lors d'un cycle de 24 heures.

$$Ip = \frac{Aj}{M_a \times N_a} = \frac{Aj}{A_m}$$

**Aj** = l'activité journalière.

**Ma** = le maximum d'activité de la journée.

**Na** = la durée de l'activité.

**Am** = activité maximale

**Ip** varie entre 0 et 1 : il tend vers 0 pour des profils présentant des pics d'activité, il tend vers 1 lorsque le profil est en plateau c'est-à-dire qu'il y a au cours de la journée une augmentation rapide vers un maximum avec une phase d'activité constante plus ou moins longue puis une décroissance rapide.

## 2- Résultats

### a- Mise en évidence des rythmes journaliers d'activité locomotrice

Dans nos conditions expérimentales, l'activité locomotrice des individus des deux espèces présente des variations cycliques (figure). On remarque que les déplacements ont lieu essentiellement pendant la phase lumineuse. Mais il y a une légère anticipation chez les individus *E. vuilleti* qui démarrent leur activité avant l'éclairage, cette anticipation est plus importante chez les mâles. Chez les deux espèces, l'activité augmente lentement pendant les toutes premières heures de la photophase puis reste constante à son maximum pendant 3 à 4 heures chez les mâles et commence aussi à décroître lentement jusqu'à l'extinction. Chez les femelles la phase d'activité maximale est encore plus courte et dure entre 2 et 3 heures maximum et la décroissance de l'activité est plus précoce puisqu'elle suit presque immédiatement l'augmentation.

### b- Variabilité des paramètres caractérisant les rythmes d'activité

#### b.1- Variabilité intra spécifique

Dans les deux conditions climatiques la quantité journalière d'activité exprimée par les femelles est plus importante que celle des mâles (tableau 1). Les taux d'activité journalière montrent que les femelles consacrent plus de temps au déplacement par rapport aux mâles. La comparaison des heures médianes d'activité montre que les mâles ont une activité journalière plus précoce puisqu'ils commencent toujours leur activité 1 heure ou plus avant les femelles (tableau 1) .

### b.2- Variabilité interspécifique

La comparaison des deux espèces a été faite chez les femelles car ce sont elles qui jouent un rôle important dans les phénomènes de compétition interspécifique. L'enregistrement de l'activité locomotrice a été fait simultanément sur 53 femelles dont 26 *E. orientalis* et 27 *E. vuilleti*. Les taux d'activité et par conséquent les indices de profil (les deux espèces ayant des profils d'activité similaires) des femelles accouplées des deux espèces ne sont pas significativement différents en conditions chaudes (Tableau 2). On observe que ce taux est différent chez les deux espèces à faible température : le temps consacré aux déplacements au cours d'une journée est plus élevé chez les femelles de *E. orientalis* (Tableau 2).

Quant au moment de démarrage de l'activité locomotrice, les femelles *E. vuilleti* sont plus précoces que les femelles *E. orientalis*. Dans les deux conditions thermiques les femelles *E. vuilleti* sont actives bien avant le début de l'éclairage (4 à 5 heures avant l'éclairage). Ainsi les heures médianes d'activité de *E. vuilleti* sont plus précoces dans toutes les conditions climatiques (tableaux 2). A 28° C les heures médianes d'activité de *E. orientalis* et *E. vuilleti* sont respectivement 12 h 14 mn ± 19 mn et 12 h 49 mn ± 10 mn, tandis qu'à 21° C ces valeurs sont respectivement 12 h 24 mn ± 24 mn et 14 h 03mn ± 24 mn. On remarque que l'écart est beaucoup plus important à basse température.

### 3- Discussion

*E. orientalis* et *E. vuilleti* sont des espèces sympatriques qui exploitent les mêmes hôtes tant dans les champs que dans les greniers qui sont d'ailleurs un milieu confiné où la compétition doit être rude. Comment expliquer la coexistence de ces deux Eupelmidae dans les stocks lorsqu'on sait que les œufs d' *E. vuilleti* arrivent difficilement au stade adulte quand ils sont déposés sur les hôtes préalablement parasités par *E. orientalis* ? A l'aide d'un système automatique d'analyse d'images, nous avons étudié l'activité locomotrice des deux espèces en tenant compte des facteurs climatiques. Comme tout appareillage de laboratoire, notre système de mesure constitue un environnement artificiel qui est susceptible d'influencer le comportement des parasitoïdes. Il permet toutefois de limiter la variabilité des conditions externes et mettre en évidence l'expression endogène de l'activité.

Les résultats que nous obtenons montrent que les deux espèces présentent des variations journalières de leur activité locomotrice comme la grande majorité des insectes [18]. L'existence des rythmes d'activité locomotrice a également été mise évidence chez d'autres hyménoptères parasitoïdes. En effet, chez les hyménoptères parasitoïdes de Drosophiles l'activité locomotrice varie de façon circadienne [1], le même phénomène a été observé chez les Trichogrammes [21]. L'étude des paramètres caractérisant l'activité locomotrice a permis une meilleure description des rythmes d'activité qui est utile notamment dans le cas des variations plus subtiles dues à certains facteurs. La variabilité de l'activité locomotrice observée à différentes échelles résulterait des caractéristiques endogènes différentes [21]. En effet, la différence entre les taux d'activité journalière des mâles et des femelles pourraient provenir de significations différentes de l'activité locomotrice : chez les femelles elle est surtout liée à la recherche du site d'oviposition [22, 23, 1] alors que chez les mâles elle est en relation avec la recherche de partenaires sexuels [24, 25].

La comparaison de l'activité des deux espèces montre qu'elles ont des taux et des profils d'activité semblables. Cependant l'heure de démarrage d'activité et l'heure médiane d'activité sont les paramètres qui différencient nettement les deux Eupelmidae. Les individus *E. vuilleti* sont capables d'émerger plusieurs heures avant le signal lumineux alors que les *E. orientalis* émergent exclusivement pendant la photophase et leur activité est très dépendante de la lumière [26]. Cette différence de sensibilité au signal photopériodique a donc pour conséquence un décalage de démarrage de l'activité au cours de la journée, ce qui permet de réduire les contacts entre les organismes des deux espèces. Dans la nature, ces Eupelmidae exploitent les mêmes hôtes, on peut de ce fait considérer que le déphasage d'activité est un moyen de réduire l'intensité de la compétition. Les études sur l'hyperparasitisme facultatif de ces parasitoïdes montrent que les oeufs de *E. vuilleti* déposés sur des hôtes parasités par *E. orientalis* ont de faibles chances d'atteindre le stade adulte [27]. Cela constitue probablement une pression qui a conduit *E. vuilleti* à démarrer plus tôt son activité journalière afin de parasiter les hôtes sains disponibles avant qu'ils soient découverts par les femelles de *E. orientalis*.

Le rôle du déphasage d'activité dans la coexistence des espèces sympatriques a été évoqué pour plusieurs types d'associations intéressant les insectes [28]. Dans le cas de ces Eupelmidae il est probable que l'activité précoce de *E. vuilleti* soit un facteur d'efficacité de la recherche d'hôtes permettant de compenser une capacité compétitive plus faible. L'étude de la capacité de recherche d'hôtes par ces Eupelmidae doit être envisagée en situation de compétition afin de mieux comprendre le rôle de l'activité précoce de *E. vuilleti*.

## **BIBLIOGRAPHIE**

**Tableau 1** : Variabilité entre sexes des paramètres caractérisant le rythme d'activité chez *E. orientalis* dans les conditions "chaudes" (28° C ± 2° C, 75 % r.h., L.D. 12 : 12) et dans les conditions "froides" (21° C ± 2° C, 75 % h. r., L.D. 12 : 12). Les valeurs présentées dans le tableau sont des moyennes ± écart-type. La comparaison des moyennes s'est faite par le test t de Student (N. S. = différence non significative).

- **Activité locomotrice de *E. orientalis***

• Conditions "chaudes".

|  | Taux d'activité journalière | Heure médiane d'activité (H.M.A.) | Indice de profil (I. P.) |
|---|---|---|---|
| -Femelles | 30,43 ± 0,70<br>n=15 | 12h58 ± 12mn<br>n=15 | 0,38 ± 0,01<br>n=15 |
| -Mâles | 14,22 ± 2,59<br>n=12 | 11h48 ± 30mn<br>n=12 | 0,53 ± 0,13<br>n=12 |
| -Valeur de t | 22,35<br>(p < 0,001) | 3,47<br>(p < 0,01) | 4,28<br>(p < 0,001) |

• Conditions "froides".

|  | Taux d'activité journalière | Heure médiane d'activité H.M.A.) | Indice de profil (I. P.) |
|---|---|---|---|
| -Femelles | 27,95 ± 1<br>n=13 | 13h25 ± 18mn<br>n=13 | 0,37 ± 0,03<br>n=13 |
| -Mâles | 24,59 ± 2,68<br>n=13 | 13h10 ± 8mn<br>n=13 | 0,36 ± 0,03<br>n=13 |
| -Valeur de t | 4,09<br>(p < 0,001) | 2,67<br>(p < 0,002)) | 0,9<br>(N. S.) |

- **Activité locomotrice de *E. vuilleti***

• Conditions "chaudes".

|  | Taux d'activité journalière | Heure médiane d'activité H.M.A.) | Indice de profil (I. P.) |
|---|---|---|---|
| -Femelles | 25,84 ± 2,07<br>n=15 | 12h21 ± 24mn<br>n=15 | 0,40 ± 0,02<br>n=15 |
| -Mâles | 13,75 ± 1,18<br>n=12 | 11h48 ± 12mn<br>n=12 | 0,48 ± 0,11<br>n=12 |
| -Valeur de t | 17,52<br>(p < 0,001) | 4,18<br>(p < 0,001) | 2,66<br>(p < 0,002) |

• Conditions "froides".

|  | Taux d'activité journalière | Heure médiane d'activité H.M.A.) | Indice de profil (I. P.) |
|---|---|---|---|
| -Femelles | 31,23 ± 3,82<br>n=13 | 11h58 ± 34mn<br>n=13 | 0,40 ± 0,02<br>n=13 |
| -Mâles | 27,22 ± 1,15<br>n=12 | 12h30 ± 17mn<br>n=12 | 0,41 ± 0,02<br>n=12 |
| -Valeur de t | 3,36<br>($p < 0,002$) | 2,81<br>($p < 0,002$) | 0,9<br>(N. S.) |

**Tableau 2** : Paramètres caractérisant le rythme d'activité chez les femelles accouplées d'*E. orientalis* et d'*E. vuilleti*. Les valeurs présentées dans le tableau sont des moyennes ± écart-type. La Comparaison entre femelles des deux espèces s'est faite pour chaque condition thermopériodique avec le test t de Student (N. S. = différence non significative).

• Conditions "chaudes".

|  | Heure médiane d'activité (H. M. A.) | Taux d'activité journalière (%) | Indice de profil (I. P.) |
|---|---|---|---|
| *E. orientalis.* | 12h49 ± 10mn<br>n=16 | 22,36±2,11<br>n=16 | 0,38±0,05<br>n=16 |
| *E. vuilleti* | 12h14±19mn<br>n=18 | 24,12±3,43<br>n=18 | 0,38±0,02<br>n=18 |
| Test t | 6,40<br>(p < 0,001) | 1,72<br>(N.S.) | 0<br>(N.S.) |

• Conditions "froides".

|  | Heure médiane d'activité (H. M. A.) | Taux d'activité journalière (%) | Indice de profil (I. P.) |
|---|---|---|---|
| *E. orientalis.* | 14h03 ± 24mn<br>n=10 | 21,25 ± 3,50<br>n=10 | 0,46 ± 0,02<br>n=10 |
| *E. vuilleti* | 12h24 ± 24mn<br>n=9 | 14,51 ± 2,28<br>n=9 | 0,34 ± 0,02<br>n=9 |
| Test t | 8,49<br>(p < 0,001) | 4,68<br>(p < 0,001) | 12,35<br>(p < 0,001) |

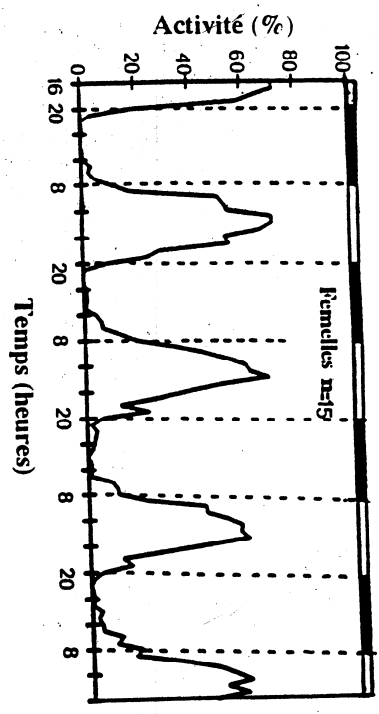
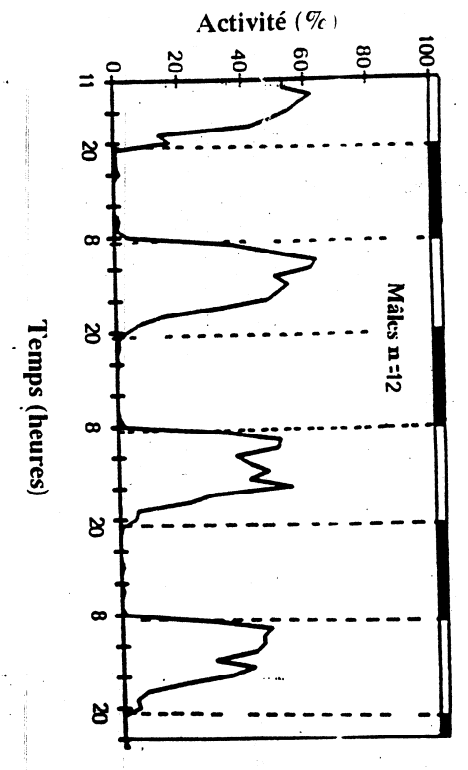
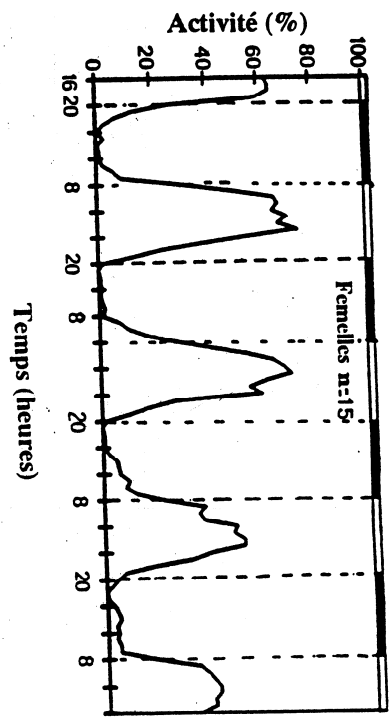
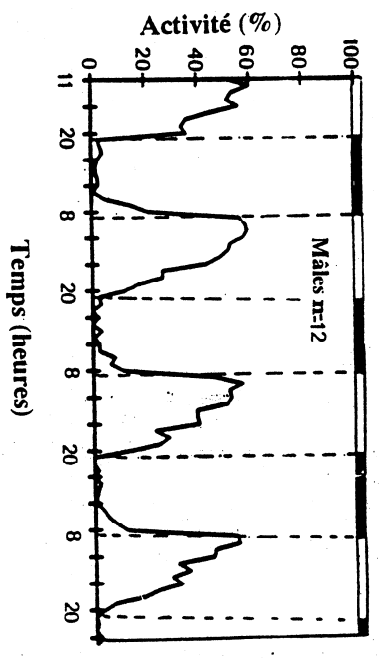

**Figure** - Mise en évidence des rythmes journaliers d'activité locomotrice en conditions " chaudes " (28° C ± 2° C, 75 % r.h., L.D. 12 : 12) chez *E. orientalis* et *E. vuilleti*.